\documentclass{llncs}
\usepackage{makeidx}
\usepackage{amssymb}

\def\FO{\textsf{FO}}
\def\calJ{\mathcal{J}}
\def\calL{\mathcal{L}}
\def\calR{\mathcal{R}}
\def\J{\textbf{J}}
\def\LL{\textbf{L}}
\def\RR{\textbf{R}}
\def\V{\textbf{V}}
\def\Box{\hbox{\rlap{$\sqcap$}$\sqcup$}}
\def\block{\mathbin{\Box}}

\begin{document}

\title{From algebra to logic: There and back again \\ The story of a hierarchy\thanks{This work was partially supported by the ANR through ANR-2010-BLAN-0204}}

\author{Pascal Weil\inst{1,}\inst{2}}
\institute{
CNRS, LaBRI, UMR 5800, F-33400 Talence, France,
\email{pascal.weil@labri.fr} \and
Univ. Bordeaux, LaBRI, UMR 5800, F-33400 Talence, France.}

\maketitle

Formal language theory historically arose from the definition of models of computation (automata, grammars, etc) and relied for its first step on combinatorial reasoning, especially combinatorics on words. Very quickly however, algebra and logic were identified as powerful tools for the classification of rational languages, e.g. with the definition of the syntactic monoid of a language and B\"uchi's theorem on monadic second-order logic. It didn't take much time after that to observe that, conversely, formal language theory is itself a tool for algebra and logic.

The results which we will present are an illustration of this back-and-forth movement between languages, algebra and logic. They deal with a hierarchy of classes of rational languages which arises in different contexts and turned out to solve a problem in logic, namely the decidability of the quantifier alternation hierarchy within the two-variable fragment of first-order logic $\FO^2[<]$.

The full picture uses a collection of results in logic, combinatorics on words and algebra which were obtained independently of the quantifier alternation hierarchy by various authors over several decades.

Let $\calR_0$ be the class of piecewise testable languages, which is natural from a combinatorial and automata-theoretic point of view, and corresponds to the first level of the quantifier alternation hierarchy within $\FO^2[<]$ (and within $\FO[<]$) as well). This class is rather simple and reasonably well understood, see \cite{1986:Pin,1994:Almeida}. We first consider the hierarchies of classes of languages obtained from $\calR_0$ by alternatingly closing it under deterministic and co-deterministic closure:  we let $\calL_0 = \calR_0$, $\calR_{k+1}$ (resp. $\calL_{k+1}$) be the deterministic (resp. co-deterministic) closure of $\calL_k$ (resp. $\calR_k$).

Results from the 1970s and 1980s \cite{1976:Schutzenberger,1980:Pin} show that the classes $\calR_k$ and $\calL_k$ are varieties (whether a language $L$ belongs to one of these classes depends only on its syntactic monoid) and describe the corresponding varieties of finite monoids $\RR_k$ and $\LL_k$. Results from the 1960s \cite{1965:KrohnRhodesTilson} (see also \cite{1993:Weil,2009:RhodesSteinberg,2010:KufleitnerWeil} shows that their membership problems are decidable and they form an infinite hierarchy.

A first view of the structure of the lattice formed by these varieties can be obtained by using purely algebraic results from the 1970s on a seemingly different hierarchy, that of varieties of idempotent monoids \cite{1970:Gerhard}. The theory of the latter varieties is particularly well understood, and one can exhibit for each of them structurally elegant identities and solutions of the word problem (of the corresponding relatively free object) \cite{1989:GerhardPetrich}.

To completely elucidate the structure of the lattice generated by the $\calR_k$ and $\calL_k$, Kufleitner and Weil introduced the notion of condensed rankers \cite{2012:Kuf-leitnerWeil}. These are a rather natural extension of the algorithm to solve the word problem in the relatively free idempotent monoids and have natural connections with deterministic and codeterministic products. But they are also -- and foremost -- a variant of the rankers introduced by Weiss and Immerman \cite{2009:WeisImmerman} (following the turtle programs of Schwentick, Th\'erien and Vollmer \cite{2001:SchwentickTherienVollmer}) to characterize the levels of the quantifier alternation hierarchy of $\FO^2[<]$. As a result one can show that the $k$-th level of this hierarchy coincides with the intersection $\calR_{k+1} \cap \calL_{k+1}$, thus proving the decidability of each level of the hierarchy \cite{2012:KufleitnerWeil-CSL}.

The story does not end there: using algebraic methods similar to those described in his book \cite{1994:Straubing}, Straubing showed \cite{2011:Straubing} that the $k$-th level of the quantifier alternation hierarchy of $\FO^2[<]$ is the variety of languages whose syntactic monoid is in the $k$-th term of the sequence given by $\V_1 = \J$ and $\V_{n+1} = \V_n \block \J$. Here $\J$ is the class of $\calJ$-trivial monoids, which characterizes piecewiste testable languages by Simon's theorem \cite{1975:Simon} and $\block$ denotes the two-sided block product, the bilateral version of the more classical wreath product. Then Straubing and Krebs showed that every class of finite monoids is decidable \cite{2012:KrebsStraubing}, thus providing an alternate proof of the decidability of the quantifier alternation hierarchy, but also giving an alternative characterization of the classes $\V_k$: a finite monoid $M$ is in $\V_k$ if and only if it sits in both $\RR_{k+1}$ and  $\LL_{k+1}$.

The coincidence of these two very differently defined hierarchies raises an intriguing question: what connects the block product with the alternate operation of deterministic and co-deterministic closure?\dots



\end{document}